\documentclass[twocolumn,secnumarabic,amssymb, nobibnotes, aps, prd,superscriptaddress]{revtex4-1}
\pdfoutput=1
\usepackage{amsmath,graphicx,hyperref,longtable,amsthm}

\newcommand\T{\rule{0pt}{2.6ex}} 
\newcommand\B{\rule[-1.2ex]{0pt}{0pt}}

\begin{document}

\title{The Impact of Assuming Flatness in the Determination \\ of Neutrino Properties from Cosmological Data}%

\author{Aaron Smith}
\email{aaronrs@byu.edu}
\affiliation{Physics Department, Brigham Young University, Provo, UT 84602}

\author{Maria Archidiacono}
\affiliation{Physics Department and INFN, Universita' di Roma ``La Sapienza'', Ple Aldo Moro 2, 00185, Rome, Italy}

\author{Asantha Cooray}
\affiliation{Center for Cosmology, Dept. of Physics \& Astronomy, University of California, Irvine, CA 92697} 

\author{Francesco De Bernardis}
\affiliation{Center for Cosmology, Dept. of Physics \& Astronomy, University of California, Irvine, CA 92697} 

\author{Alessandro Melchiorri}
\affiliation{Physics Department and INFN, Universita' di Roma ``La Sapienza'', Ple Aldo Moro 2, 00185, Rome, Italy}

\author{Joseph Smidt}
\affiliation{Center for Cosmology, Dept. of Physics \& Astronomy, University of California, Irvine, CA 92697}

\date{\today}%

\begin{abstract} 
Cosmological data have provided new constraints on the number of neutrino species and the neutrino mass. However these constraints depend on assumptions related to the underlying cosmology. Since a correlation is expected between the number of effective neutrinos $N_{eff}$, the neutrino mass $\sum m_\nu$, and the curvature of the universe $\Omega_k$, it is useful to investigate the current constraints in the framework of a non-flat universe. In this paper we update the constraints on neutrino parameters by making use of the latest cosmic microwave background (CMB) data from the ACT and SPT experiments and consider the possibility of a universe with non-zero curvature. We first place new constraints on $N_{eff}$ and $\Omega_k$, with $N_{eff} = 4.03 \pm 0.45$ and $10^3 \, \Omega_k = -4.46 \pm 5.24$. Thus, even when $\Omega_k$ is allowed to vary, $N_{eff} = 3$ is still disfavored with 95\% confidence. We then investigate the correlation between neutrino mass and curvature that shifts the $95 \%$ upper limit of  $\sum m_\nu < 0.45$ eV to $\sum m_\nu < 0.95$ eV. Thus, the impact of assuming flatness in neutrino cosmology is significant and an essential consideration with future experiments. 

\end{abstract}

\maketitle

\section{Introduction}

Throughout the previous decades experimental cosmology has benefited from accurate measurements of the cosmic microwave background (CMB). The data have determined constraints on several cosmological parameters to remarkable accuracy and the ability to constrain new physics with the CMB continues to improve. Future CMB experiments might even be able to measure B-mode polarization and distinguish between neutrino hierarchy models. However, when constraining new parameters one must be careful when constraints depend on assumptions about the underlying cosmology. For example, a correlation between the neutrino properties and the curvature of the universe is clearly expected since, a higher number of neutrino species  or large mass would introduce pre-recombination effects, shifting the positions of the peaks in the angular CMB spectrum (cf.~\cite{cormelk,Mantz:2009rj}).

Here we present an update on the constraints of the number of neutrino species $N_{eff}$ and the sum of neutrino masses $\Sigma m_{\nu}$ in the framework of non-flat universes with  $\Omega_k \neq 0$  combining the Wilkinson Microwave Anisotropy Probe (WMAP) 7-year~\cite{Komatsu:2010fb}, South Pole Telescope (SPT)~\cite{Keisler:2011aw} and Atacama Cosmology Telescope (ACT)~\cite{Dunkley:2010ge} datasets. 



The paper is organized as follows. In Section~\ref{sec:theory} we 
give theoretical arguments for why the $N_{eff}$ and $\Omega_k$ 
parameters should be correlated. In Section~\ref{sec:method} we discuss our method of constraining the parameters $N_{eff}$, $\Sigma m_{\nu}$, and $\Omega_k$. We present the results of the analysis in Section~\ref{sec:imp}. Finally, in Section~\ref{sec:conclusion} we conclude and discuss the implications of assuming flatness in neutrino cosmology. 

\section{The Effective Neutrino Number}
\label{sec:theory}
The effective neutrino number $N_{eff}$ is defined as the contribution of neutrinos to the relativistic degrees of freedom $g_*$. In a standard physics scenario the particles contributing to the total value of $g_* \simeq 10.75$ are electrons, three neutrinos (and their antiparticles), and photons. Any extra relativistic degrees of freedom can be parameterized in terms of an excess with respect to the standard effective neutrino number $N_{eff}=3$ (which more precisely is $\simeq3.046$ after accounting for QED corrections and non-instantaneous decoupling of neutrinos)~\cite{Hannestad:2005ey,Calabrese:2011hg}. The neutrino energy density is:
\begin{equation}
\rho_\nu = N_{eff} \, \frac{7}{8} \, \left(\frac{4}{11}\right)^{4/3} \rho_\gamma \, ,
\end{equation}
where $\rho_\gamma$ is the energy density of photons. A first effect of $N_{eff}$ is related to the primordial helium abundance $Y_P$. Changing $N_{eff}$ affects the freeze-out temperature $T_{freeze}$ during BBN and therefore the final neutron to proton ratio $n_n/n_p$~\cite{Hamann:2011ge}. Larger $N_{eff}$ means earlier freeze-out, larger $n_n/n_p$, and larger $Y_p$. 

The effect of $N_{eff}$ on cosmological observables (e.g. CMB anisotropy power spectrum and galaxy power spectrum) is emphasized by 
the epoch of matter-radiation equality $a_{eq}$. In particular, for what concern the CMB, an increase in $a_{eq}$ changes the extent of the early Integrated Sachs-Wolfe effect. The relation between $a_{eq}$ and $N_{eff}$ is given by equating energy densities:
\begin{equation}
\rho_{rad} = \rho_m \quad \Longleftrightarrow \quad a_{eq} = \frac{1 + 0.227 N_{eff}}{40484 \, \Omega_m \, h^2} \, .
\end{equation}
This shows a linear relationship for $a_{eq}(N_{eff})$, which transfers to the baryon to photon ratio at equality~\cite{Dodelson:2003ft}:
\begin{align} 
R_{eq} & = \frac{3 \rho_b}{4 \rho_\gamma} \Big|_{a_{eq}} \, , \notag \\
& = 30496 \, \Omega_b \, h^2 \, a \, \Big|_{a_{eq}} \, , \notag \\
& = \frac{1 + 0.227 N_{eff}}{1.3276} \frac{\Omega_b}{\Omega_m} \, . 
\end{align}
The presence of baryons in the relativistic cosmic fluid slows down the sound speed according to the definition, \begin{equation} c_s \equiv 1/\sqrt{3(1+R)} \, , \end{equation} and so this quantity is also affected at equality by the effective neutrino number. This reflects in the size of sound horizon at a generic time $\tau$~\cite{Dodelson:2003ft}: 
\begin{align}
\label{eqn:rsn}
r_s & \equiv \int_0^\tau \, d\tau' \, c_s (\tau') \, , \notag \\ 
& = \int_0^a \, \frac{da}{a^2 H} \, c_s(a) \, , \notag \\
& \approx \frac{2}{3 k_{eq}} \, \sqrt{\frac{6}{R_{eq}}} \, \ln\left\{\frac{\sqrt{1+R} \, + \sqrt{R+R_{eq}}}{1\, +\sqrt{R_{eq}}} \right\} \, , \notag \\
& = \frac{6.612 \times 10^{-3}}{H_0 \sqrt{\Omega_m \Omega_b h^2}} \, \ln\left\{\frac{\sqrt{1+R} \, + \sqrt{R+R_{eq}}}{1\, +\sqrt{R_{eq}}} \right\} \, .
\end{align}
The last equations come from assuming the Universe is matter dominated during recombination. As can be seen, the sound horizon depends on $N_{eff}$ through $R_{eq}$. 

\begin{figure}[h]
\centering
\includegraphics[width=3.25in]{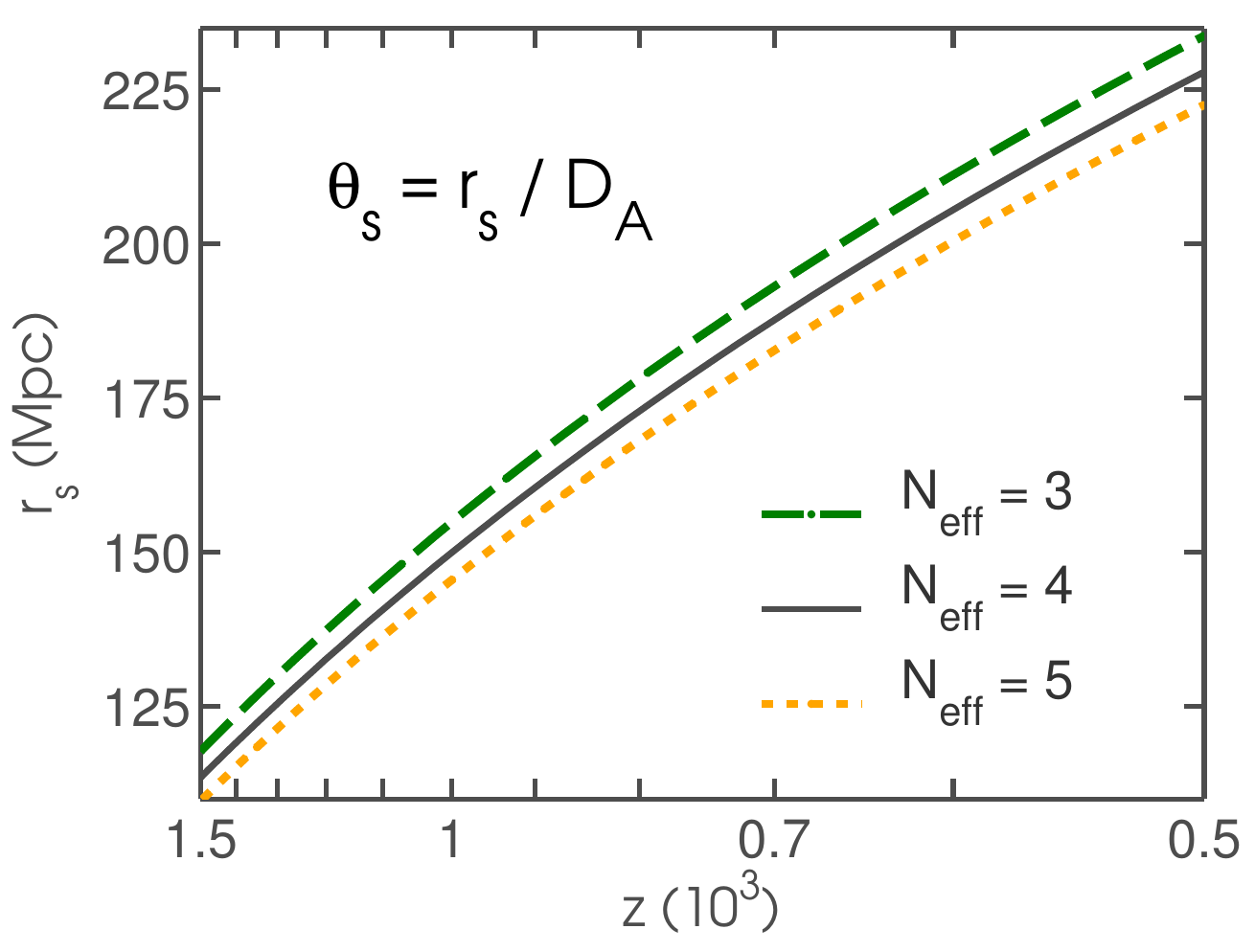} 
\vspace{-.1cm}
\caption{\label{fig:rsneff}A demonstration of how the sound horizon $r_s$ changes with the effective neutrino number $N_{eff}$ under the matter dominated approximation given by Eq.~\ref{eqn:rsn}.}
\end{figure}

\begin{table*} 
 \label{tab:one}

 \begin{tabular}{@{} c cccc  @{}}
 \hline \hline
Parameter \T \B& WMAP7+BAO+$H_0$ \quad & WMAP7+SPT & \quad WMAP7+ACT & WMAP7+SPT+ACT \,\, \\
 \hline
$100 \Omega_bh^2$ \T & $2.249 \pm 0.054$ & $2.256 \pm 0.041$ & $2.235 \pm 0.047$ & $2.258 \pm 0.040$ \\
$\Omega_ch^2$ & $0.135 \pm 0.016$ & $0.130 \pm 0.0094$ & $0.137 \pm 0.012$ & $0.129 \pm 0.0091$ \\
$\Omega_\Lambda$ & $0.721 \pm 0.018$ & $0.722 \pm 0.015$ & $0.714 \pm 0.018$ & $0.722 \pm 0.015$ \\
$n_s$ & $0.979 \pm 0.015$ & $0.9808 \pm 0.0122$ & $0.982 \pm 0.013$ & $0.9803 \pm 0.0121$ \\
$\tau$ & $0.086 \pm 0.014$ & $0.085 \pm 0.014$ & $0.086 \pm 0.014$ & $0.086 \pm 0.014$ \\
\,\, $H_0$ (km/s/Mpc) \B & $75.1 \pm 3.4$ & $74.0 \pm 2.0$ & $74.6 \pm 2.15$ & $73.9 \pm1.92$ \\
\hline
$N_{eff}$ \T \B & $4.34 \pm 0.88$ & $3.91 \pm 0.43$ & $4.30 \pm 0.58$ & $3.89 \pm 0.41$ \\
 \hline \hline
\end{tabular}
\caption{\label{tab:wmap7} Summary of matching results from WMAP 7-year~\cite{Komatsu:2010fb}, SPT~\cite{Keisler:2011aw}, and ACT~\cite{Dunkley:2010ge}. Note that the analyses are modeled by the choice to reproduce SPT results, which produces a smaller value for $N_{eff}$ than expected for ACT data. All datasets include BAO and $H_0$ for improved parameter constraints. The quoted errors are given at the 68\% confidence levels (CL).} 
\end{table*}

In recent papers~(see e.g.~\cite{Hou:2011ec,archi01,giusarma11}) it has been found that the number of neutrinos is greater than the standard model value at more than $2\sigma$. The presence of additional neutrinos can be described by a (3+1) or (3+2) model with three active neutrinos and one or two sterile neutrinos~\cite{Hamann:2010bk,Giunti:2011gz}. More exotic solutions may include arguments supporting modified dark energy models~\cite{Volkas:2001zb,Shaposhnikov:2007nf,Sanders:2007zn}. In this work we 
explore possible overlooked parameter degeneracies that could still favor the standard model without the introduction of new physics.


In~\cite{Hou:2011ec} the authors provide qualitative arguments for how changing the number of allowed neutrinos affects the observed values of parameters. One example which we use is the relative dependence of distance measurements on the Hubble constant. In fact, the sound horizon at recombination scales as $r_s \propto 1/H$ while the distance a photon typically diffuses prior to its last scattering goes as $r_d \propto 1/\sqrt{H}$. This is significant because the response of the radiation relative to matter determines the degree of damping prior to recombination. In other words, with $\theta_s = r_s/D_A$ fixed by observation, the angular diameter distance, $D_A$, must also decrease as $1/H$ which is more rapid than $r_d$. Thus, the damping increases according to $\theta_d = r_d/D_A \propto \sqrt{H}$~\cite{Hou:2011ec}. If these distances vary according to $H$ then they also vary according to any parameter correlated with $H$. In 
an open universe with nonzero curvature the effective neutrino number is slightly reduced. The theory confirms this because as stated above $\theta_s$ is constrained by observation which means if $N_{eff}$ is reduced and $\Omega_k > 0$ then $r_s$ and $D_A$ both increase (see Fig.~\ref{fig:rsneff}). However, if the parameter space favors a closed universe then there will appear to be a higher number of effective neutrinos. This is one of the primary reasons for expecting correlation between $N_{eff}$ and $\Omega_k$. 

We conclude this section by stating our purpose 
further constrain the neutrino mass and investigate the effect of curvature on this parameter. 
Finally, we acknowledge that there are many papers on the subject of constraining neutrino parameters. A non-exhaustive list of additional references 
includes Refs.~\cite{Agarwal:2010mt,DeBernardis:2009di,Jimenez:2010ev,Hannestad:2003xv,Hannestad:2010yi,Cirelli:2006kt,Zunckel:2006mt,Seljak:2006bg,Acero:2008rh,DeBernardis:2008qq,Tereno:2008mm,Gong:2008pg,Reid:2009nq,Shimon:2010gs}. 

\section{Analysis Method}
\label{sec:method}

In order to fit cosmological models to data we use a modified version of the publicly available CosmoMC software package~\cite{Lewis:2002ah}. This uses a Monte Carlo Markov Chain analysis on calculations of the lensed CMB power spectrum made with the CAMB package. Our 
analysis combines 
the following CMB anisotropy datasets: WMAP 7-year~\cite{Komatsu:2010fb}, SPT~\cite{Keisler:2011aw}, and ACT~\cite{Dunkley:2010ge}. Including BAO+$H_0$ simply means we are using the baryon acoustic oscillation (BAO) data of Percival \textit{et al.}~\cite{Percival:2009xn} and impose a prior on the Hubble parameter based on the last Hubble Space Telescope observations~\cite{hst}. We integrate spectral data out to $\ell_{\rm max}=3000$. We sample from the following parameters: the baryon~$\Omega_bh^2$, cold dark matter~$\Omega_ch^2$, and dark~$\Omega_\Lambda$ energy densities, 
the scalar spectral index~$n_s$, the optical depth to reionization~$\tau$, the Hubble parameter~$H_0$, and the amplitude of SZ spectrum~$A_{SZ}$. We also consider the effective neutrino number~$N_{eff}$, spatial curvature~$\Omega_k$, and the sum of neutrino masses~$\sum m_\nu$. 

Finally, we make decisions specific to the high multipole mode data. We consider purely adiabatic initial conditions. When the background data are taken to small enough scales the spectra from infrared source emission must be taken into account. The IR spectra is dominated by Poisson power partially from source emission clustering at the smallest scales. Thus, a model for such effects must be subtracted out from the CMB power spectra. The resulting adaptations are representative of the considerations made during the process of checking the code for consistency with established results. It is also important to include a Big Bang nucleosynthesis (BBN) consistency check during the sampling in order to provide analysis consistent with helium abundance measurements, as proposed in Refs.~\cite{Ichikawa:2006dt,Hamann:2007sb}. We remark that the ACT collaboration did not include the same BBN consistency condition used by SPT and our analysis. This explains why we find a slightly better constraint on $N_{eff}$ than Ref.~\cite{Dunkley:2010ge}. 

Additionally, we also constrain the sum of the neutrino masses  $\Sigma m_{\nu}$. To do this, we use a top hat prior on the fractional contribution of neutrinos to the total mass density, $f_\nu \equiv \Omega_\nu/\Omega_m \in [0, 0.5]$. Then we extract $\sum m_\nu$ from $f_\nu$ through the standard relation,
\begin{equation}
\label{eqn:fnu}
\sum m_\nu = 94 \Omega_\nu h^2 \text{ eV} = 94 h^2 \Omega_m f_\nu \text{ eV} \, ,
\end{equation}
where $\Omega_\nu \equiv \rho^0_\nu/\rho_{cr}$ is the neutrino contribution to the energy density. 

\section{Results}
\label{sec:imp}
Under the flat Universe scenario the constraint improves to $N_{eff} = 3.89 \pm 0.41$ at the 68\% confidence level (see Table~\ref{tab:wmap7}). 
This result suggests $N_{eff} = 3$ is inconsistent with the data with $\sim$ 95\% confidence.  

\begin{figure}
\centering
\includegraphics[width=3.3in]{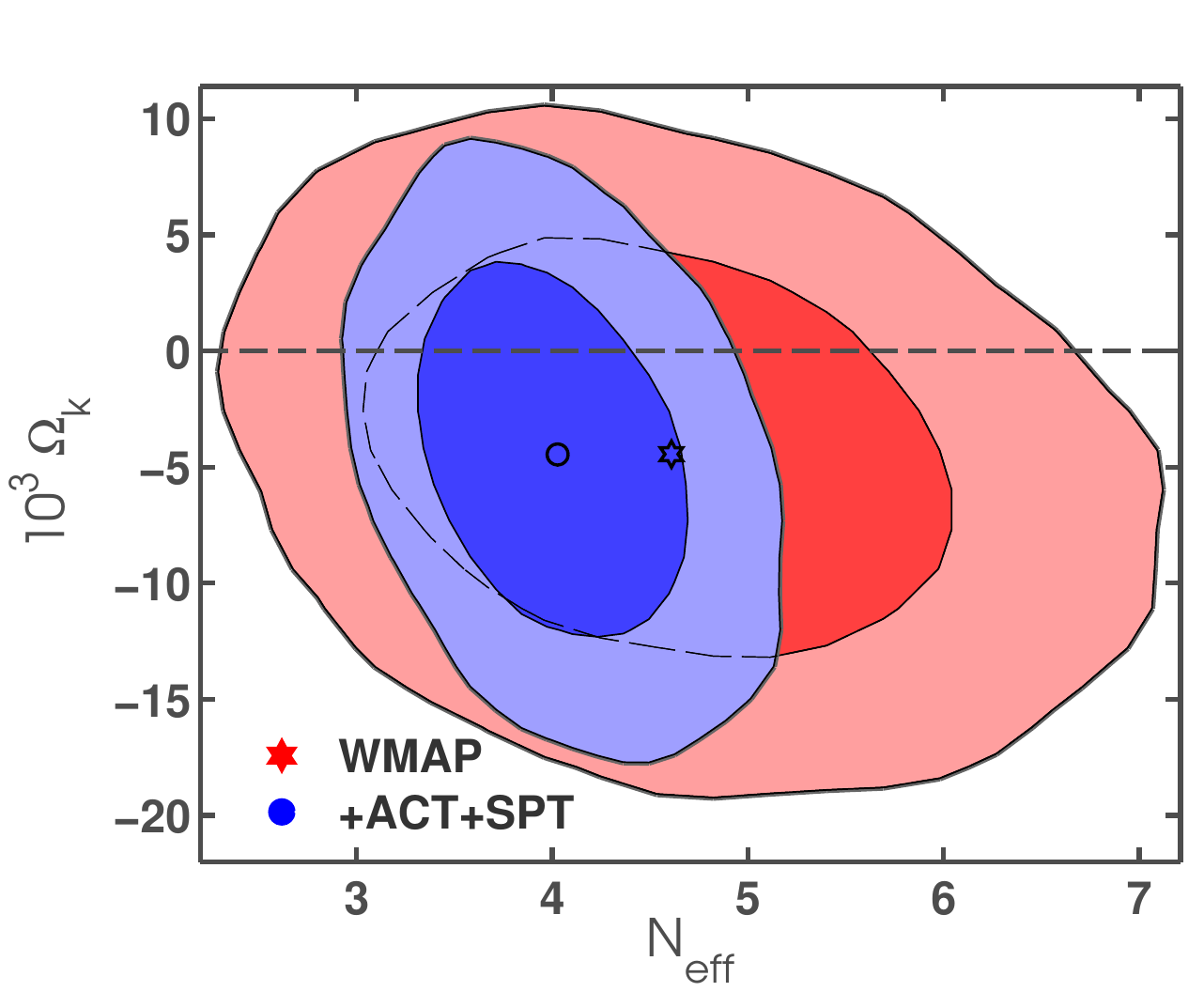}
\caption{\label{fig:omkvneff}Correlation between $\Omega_k$ and $N_{eff}$. The credible intervals are given at the 68\% and 95\% confidence levels and the markers indicate the locations of the marginalized values. WMAP+BAO+$H_0$ is shown in red while WMAP+ACT+SPT+BAO+$H_0$ is in blue. Note that the effect of adding additional datasets is significant.}
\end{figure}

We then allow the curvature to vary, to determine 
how assuming flatness affects the constraints on $N_{eff}$. Fig.~\ref{fig:omkvneff} demonstrates the correlation between $\Omega_k$ and $N_{eff}$, which agrees with the prediction from Section~\ref{sec:theory}. Interestingly, the effect of the additional CMB datasets (ACT and SPT) 
increases the correlation between these parameters with respect to WMAP 7-year data alone. This may be due in part to the considerable improvement in $N_{eff}$ whereas the uncertainty in the curvature 
is not noticeably improved by the addition of small scale anisotropy measurements. These results suggest that an open universe with fewer neutrinos would look similar to a flat universe with more neutrinos. We also note that when including $N_{eff}$ as a free parameter in the $\Lambda$CDM+$\Omega_k$ model, the $1\sigma$ constraint of $\Omega_k = -0.0023^{+0.0054}_{-0.0056}$ found in Ref.~\cite{Komatsu:2010fb} does not deteriorate significantly for the same combination of datasets (i.e. WMAP+BAO+$H_0$). This is due to the presence of the BAO data and the $H_0$ prior in the analysis, since both probes are sensitive to the geometry of the Universe. Therefore, BAO and $H_0$ help to break the degeneracy between $N_{eff}$ and $\Omega_k$.

\begin{figure}
\centering
\includegraphics[width=3.3in]{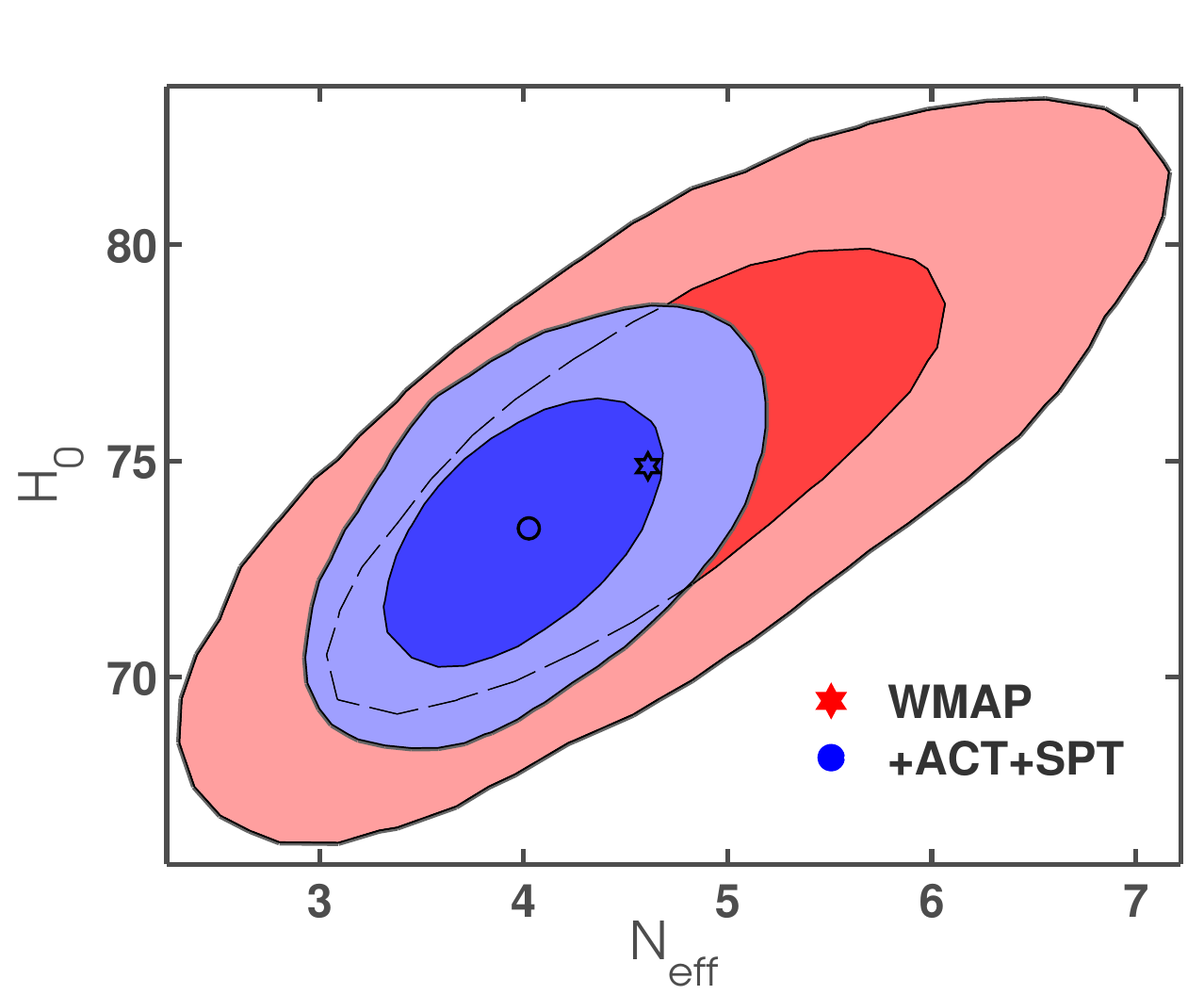}
\caption{\label{fig:h0vneff}Correlation between $H_0$ and $N_{eff}$. The credible intervals are given at the 68\% and 95\% confidence levels and the markers indicate the locations of the marginalized values. WMAP+BAO+$H_0$ is shown in red while WMAP+ACT+SPT+BAO+$H_0$ is in blue. In this case the effect of adding additional datasets is also significant.}
\end{figure}

Table~\ref{tab:omkvneff} provides a summary of parameter values for runs where $\Omega_k$ and $N_{eff}$ vary. Here we find $N_{eff} = 4.03 \pm 0.45$  and $10^3 \Omega_k = -4.46 \pm 5.24$ at the 68\% confidence level. Therefore, even when $\Omega_k$ is allowed to vary, $N_{eff} = 3$ is still disfavored with $\sim$ 95\% confidence.  Note that the increased value for $H_0$ is indicative of the known correlation between $H_0$ and $\Omega_k$. We provide an equivalent marginalized contour plot of $H_0$~vs.~$N_{eff}$ to emphasize the connection (see Fig.~\ref{fig:h0vneff}). 

\begin{table}[t]
 \begin{tabular}{@{} c cc  @{}}
 \hline \hline
Parameter \T \B & WMAP7+$N_{eff}$+$\Omega_k$ & $\quad \ldots$+ACT+SPT \,\, \\
 \hline
$100 \Omega_bh^2$ \T & $2.26 \pm 0.056$ & $2.27 \pm 0.045$ \\
$\Omega_ch^2$ & $0.136 \pm 0.0169$ & $0.129 \pm 0.00915$  \\
$\Omega_\Lambda$ & $0.721 \pm 0.0179$ & $0.723 \pm 0.0158$  \\
$n_s$ & $0.9837 \pm 0.0157$ & $0.9863 \pm 0.0147$  \\
$\tau$ & $0.0887 \pm 0.0148$ & $0.0894 \pm 0.0149$  \\
$H_0$ (km/s/Mpc) \B & $74.88 \pm 3.40$ & $73.44 \pm 2.03$  \\
\hline 
$N_{eff}$ \T & $4.61 \pm 0.96$ & $4.03 \pm 0.45$  \\
$10^3 \, \Omega_k$ \B & $-4.45 \pm 5.85$ & $-4.46 \pm 5.24$  \\
 \hline \hline 
\end{tabular}
\caption{\label{tab:omkvneff} Summary of constraints while varying $\Omega_k$ and $N_{eff}$. All datasets include BAO and $H_0$ for improved parameter 
constraints. Errors are at the 68\% CL. See Fig.~\ref{fig:omkvneff}.}
\end{table}

We now turn to the question of how well the datasets are able to constrain $\sum m_\nu$. 
Table~\ref{tab:summnu} shows the results from WMAP in the first column and the result of adding the additional datasets in the final column. Although the constraint greatly improves the two sigma limit for the masses, this is not enough to favor either the standard or inverted hierarchy. However, this is not a surprise because none of the datasets are sensitive enough on their own. Forthcoming data from the Planck experiment and other future experiments will likely improve the mass constraint~\cite{Ade:2011ah}. 

\begin{table}[t]
 \begin{tabular}{@{} c cc  @{}}
 \hline \hline
\, Parameter \T \B & WMAP7+BAO+$H_0$ & $\quad \ldots$+ACT+SPT \,\, \\
 \hline
$100 \Omega_bh^2$ \T & $2.26 \pm 0.053$ & $2.23 \pm 0.038$ \\
$\Omega_ch^2$ & $0.112 \pm 0.0036$ & $0.111 \pm 0.0029$  \\
$\Omega_\Lambda$ & $0.719 \pm 0.0182$ & $0.726 \pm 0.0154$  \\
$n_s$ & $0.968 \pm 0.0124$ & $0.963 \pm 0.0092$  \\
$\tau$ & $0.0897 \pm 0.015$ & $0.0873 \pm 0.014$  \\
$H_0$ (km/s/Mpc) \B \B & $69.2 \pm 1.6$ & $69.9 \pm 1.37$  \\
\hline
$\sum m_\nu$ \T \B & $<0.57$  eV & $<0.45$ eV \\
 \hline \hline
\end{tabular}
\caption{\label{tab:summnu} Summary of the constraint on the sum of the neutrino masses. All datasets include BAO and $H_0$ for improved parameter constraints. Errors are at the 68\% CL except for $\sum m_\nu$, which is quoted as a 95\% upper limit.} 
\end{table}

\begin{figure}[h]
\centering
\includegraphics[width=3.3in]{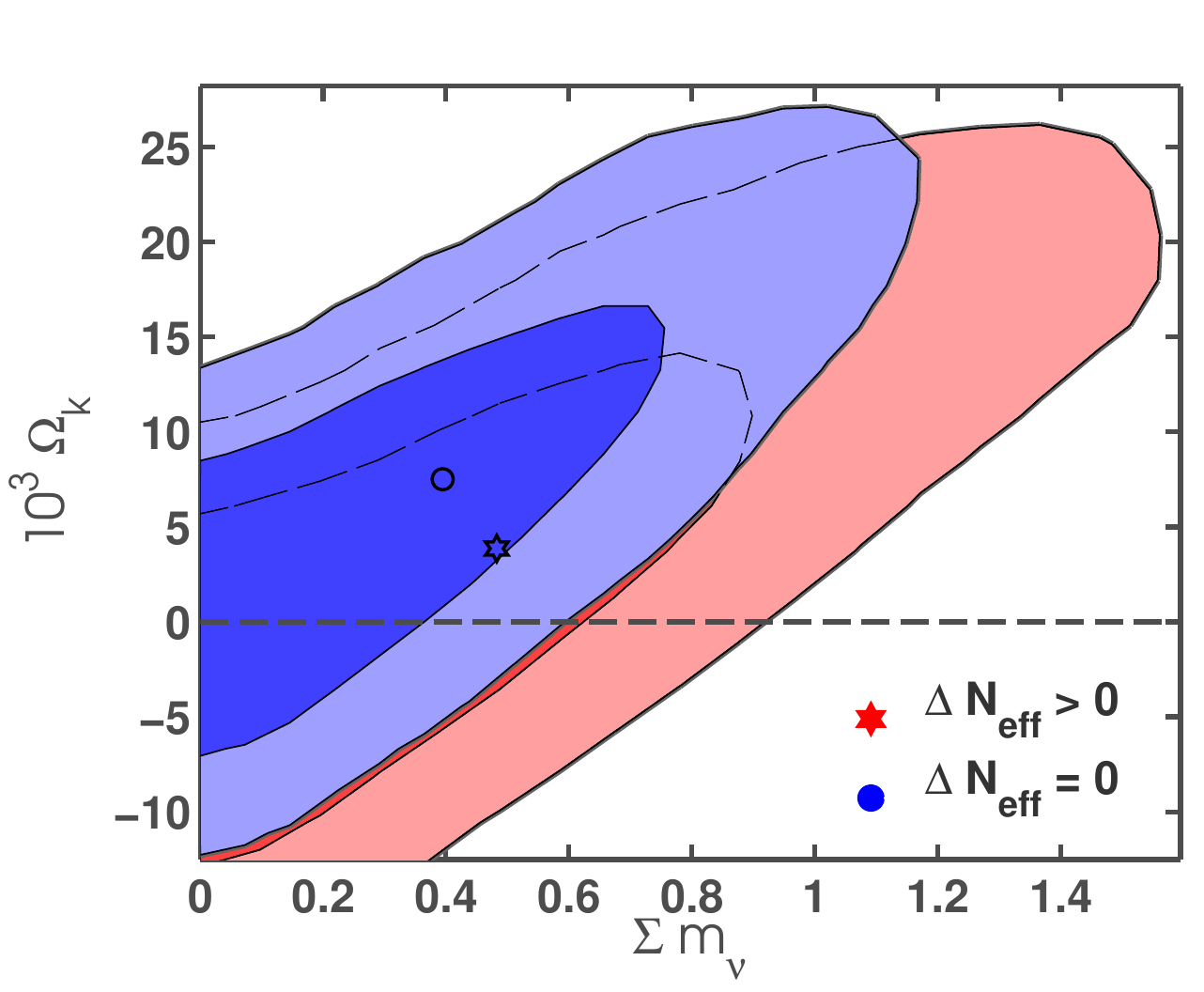}
\caption{\label{fig:omkvmass}Comparison of the correlation between $\Omega_k$ and $\sum m_\nu$ under the two $\Delta N_{eff}$ models. 
The model with three massive neutrinos is shown in blue while the model with additional relativistic species is in red. Intervals are given at the 68\% and 95\% confidence levels and markers indicate the locations of the marginalized values. Datasets include WMAP7+ACT+SPT+BAO+$H_0$. The addition of curvature allows $\sum m_\nu$ to be more than twice 
the previous constraint.} 
\end{figure}

Finally, we investigate the effect of assuming flatness while determining an upper bound on $\sum m_\nu$. 
We investigate two models. The first assumes three degenerate massive neutrinos, while the second 
allows for additional relativistic species accounted by $\Delta N_{eff} > 0$. We define the correlation coefficient $\rho_{ij}$ as the ratio of the off-diagonal term of the covariance matrix $\sigma_{ij}$ to the $1\sigma$ errors $\sigma_i \sigma_j$, so that for two parameters denoted by $i$ and $j$ we have $\rho_{ij} = \sigma_{ij}/\sigma_i\sigma_j$. Figure~\ref{fig:omkvmass} shows that $\sum m_\nu$ and $\Omega_k$ are strongly correlated with a correlation coefficient of 
$\rho_{\Omega_k \sum m_\nu} = 0.78$ for both models ($\Delta N_{eff} = 0$ and $\Delta N_{eff} > 0$). Furthermore, the degeneracy considerably increases the uncertainty in the sum of the neutrino masses. In fact, with $\Omega_k \neq 0$ the $95 \%$ upper limit on $\sum m_\nu$ more than doubles with respect to the flat case: with $\sum m_\nu < 0.95$ eV for the model assuming only three 
massive neutrinos and $\sum m_\nu < 1.19$ eV for $\Delta N_{eff} > 0$. The strong correlation between curvature and mass is expected because massive neutrinos with $m_\nu < 0.3$ eV are still relativistic until recombination so they act as an additional radiative component. As a consequence the presence of such massive neutrinos shifts the time of matter-radiation equality $a_{eq}$. Recall the discussion in Section~\ref{sec:theory} where in this case lower mass neutrinos roughly correspond to higher $N_{eff}$. Neutrinos also leave an imprint on the CMB through the early Integrated Sachs-Wolfe effect (c.f. Refs.~\cite{Lesgourgues:2006nd,Komatsu:2008hk}) which changes the position of acoustic peaks. This effect can be compensated for by a change in the geometry of the Universe, which weakens the constraints on both $\sum m_{\nu}$ and $\Omega_k$. See Table~\ref{tab:summnu2} for a summary of cosmological parameters when curvature and massive neutrinos are considered. 


\begin{table}
 \begin{tabular}{@{} c cc  @{}}
 \hline \hline
\, Parameter \T \B & $\Delta N_{eff} = 0$ & $\Delta N_{eff} > 0$ \, \\
 \hline
$100 \Omega_bh^2$ \T & $2.24 \pm 0.043$ & $2.26 \pm 0.049$ \\
$\Omega_ch^2$ & $0.118 \pm 0.0063$ & $0.134 \pm 0.0105$  \\
$\Omega_\Lambda$ & $0.711 \pm 0.0216$ & $0.703 \pm 0.0239$  \\
$n_s$ & $0.967 \pm 0.011$ & $0.982 \pm 0.015$  \\
$\tau$ & $0.0864 \pm 0.0144$ & $0.0890 \pm 0.0145$  \\
$H_0$ (km/s/Mpc) \B & $70.6 \pm 1.62$ & $73.1 \pm 2.03$  \\
\hline
$10^3 \, \Omega_k$ \T & $7.52 \pm 7.74$ & $3.46 \pm 8.69$ \\
$\sum m_\nu$ & $<0.95$ eV & $<1.19$ eV \\
$\Delta \, N_{eff}$ \B & $0$ & $0.995 \pm 0.430$ \\
 \hline \hline
\end{tabular}
\caption{\label{tab:summnu2} Summary of the constraint on the sum of the neutrino masses when $\Omega_k \neq 0$. 
$\Delta N_{eff}$ is an additional relativistic contribution after considering 3.046 massive neutrinos. Datasets include WMAP7+ACT+SPT+BAO+$H_0$. Errors are at the 68\% CL except for $\sum m_\nu$, which is quoted as a 95\% upper limit.}
\end{table}

\section{Conclusion}
\label{sec:conclusion}

The resolution of the high effective neutrino number in cosmology remains an open question. However, additional neutrinos 
may be due to parameter degeneracy or other issues in statistical analysis rather than new physics. 
The focus of this paper has been an argument for correlation between the number of effective neutrinos $N_{eff}$ and the curvature of the Universe $\Omega_k$, which arises from the effect of these parameters on distance measurements. The qualitative argument is confirmed by a statistical analysis of CMB anisotropy measurements using CosmoMC. 

In this paper we have shown that there is a correlation between $N_{eff}$ and $\Omega_k$ that gets stronger when SPT and ACT datasets are added to WMAP alone.  However, even when $\Omega_k$ is 
allowed to vary, $N_{eff} = 3$ is still disfavored by the data with 95\% confidence. Although the correlation favors a closed universe with $\Omega_k < 0$, 
if CMB data were to favor open models then the neutrino number would decrease as predicted. Perhaps 
the same element of the data that favors a closed universe may also be responsible for the trend toward a higher $N_{eff}$. More importantly, 
we find a strong correlation between curvature and 
the sum of the neutrino masses. 

Future experiments will provide further insight into both $N_{eff}$ and $\sum m_\nu$~\cite{Carbone:2010ik}. Our results are consistent with the current understanding of the data available. The strongest constraints on these parameters from the statistical analysis assuming a flat universe are $N_{eff} = 3.89 \pm 0.41$ and $\sum m_\nu < 0.45$ eV with 95\% confidence level using WMAP7+ACT+SPT+BAO+$H_0$. The constraints are weakened by 
degeneracy with the curvature parameter $\Omega_k$. However, this still represents the continued effort toward significant improvements on parameter constraints in cosmology. Although the sum of the neutrino masses is significantly improved from the WMAP 7-year result of $\sum m_\nu < 0.57$ eV, the constraint is far from being sensitive enough to rule out one of the mass hierarchies. Furthermore, we have shown that the mass uncertainty more than doubles when $\Omega_k \neq 0$. Based on our results and the estimated quality of data for Planck and other experiments, it should be possible to determine the existence or nonexistence of sterile radiation to much greater confidence in the near future. 

\section{Acknowledgments}

\noindent AS thanks members of the research group at the Center for Cosmology at UC Irvine for the stimulating environment. AS thanks Eric Hirschmann for mentoring at BYU and David Neilsen for similar help. This research was supported in part by NSF CAREER AST-0645427 (to AC) and the Department of Physics at BYU.

\end{document}